\documentclass{iopart}
\usepackage{iopams}
\usepackage{amssymb}
\usepackage{bm}
\usepackage{graphicx}

\begin{document}

\title{Quantum fluctuations of geometry in hot Universe}
\author{Iwo Bialynicki-Birula}
\address{Center for Theoretical Physics, Polish Academy of Sciences\\
Aleja Lotnik\'ow 32/46, 02-668 Warsaw, Poland}
\ead{birula@cft.edu.pl}
\begin{abstract}
The fluctuations of spacetime geometries at finite temperature are evaluated within the linearized theory of gravity. These fluctuations are described by the probability distribution of various configurations of the gravitational field. The field configurations are described by the linearized Riemann-Weyl tensor without any reference to the metric. The probability distribution of various con\-figurations is described by the Wigner functional of the gravitational field. It has a foam-like structure; dominant configurations are those with large changes of geometry at nearby points. In the high-temperature limit one obtains the Bolzmann distribution that enables one to identify the expression for the total energy of the gravitational field. The appearance of the same expression for the total energy when the gravitational field is treated as a collection of gravitons and as the high-temperature limit of the Wigner functional proves the consistency of the whole procedure. Striking differences are found between the fluctuations of the electromagnetic field and the gravitational field; among them is the divergence in the gravitational case of the probability distribution at zero temperature. This divergence is of the ``infrared type'' because it occurs in integrals over the wave vector at small $k$.
\end{abstract}
\submitto{Classical and Quantum Gravity}
\maketitle
\section{Introduction}

The main goal of this paper is to describe the quantum fluctuations of the gravitational field at finite temperature in the linearized gravity. This is done with the use of the Wigner function; a tool that has never been used in this context before. In order to reach this goal I shall introduce the quantization procedure that has the advantage of involving only the geometrical, gauge invariant elements. This is accomplished in the framework of the spinorial formulation providing another evidence for the special role played by spinors and justifying once again the praise of spinors expressed by Andrzej Trautman \cite{amt}: ``One is tempted to say that the world around us, and life in particular, are so rich because Nature found it convenient to use, among its building blocks, entities requiring spinors in their description.''

The original Wigner function is a function of positions and momenta of quantum particles. However, there is natural generalization of this concept. Namely, we may replace the canonically conjugate particle variables by their field-theoretic counterparts. This was done for the scalar field in \cite{mm} and for the electromagnetic field in \cite{ibb}. Upon this generalization, the Wigner function becomes the {\em Wigner functional} whose arguments are the field configurations.

Quantum properties of fields manifest themselves, in particular, in the field fluctuations present even in the vacuum state. In quantum electrodynamics these fluctuations lead to observable effects (Welton explanation of the Lamb shift, Casimir effect, photon shot noise). In classical physics statistical field fluctuations are described by the probability distribution of various field configurations. In quantum physics this simple description fails, due to the uncertainty relations. However, when the Wigner function is positive it may serve as a very good substitute for the standard classical distribution function.

Statistical properties of the vacuum fluctuations of the gravitational field are embodied in the probability distribution that assigns relative weights to different geometries. Exact solution of this problem is a hopeless task since it would require full-fledged quantum theory of gravity. However, an approximate solution can be obtained in linearized gravity---after all, our Universe is mostly flat. In this approach I assume that the linearized gravitational field can be quantized just like any other free field. Every {\em free field} can be viewed as a collection of uncoupled harmonic oscillators. Since the Wigner function for the thermal state of harmonic oscillators is Gaussian, it can serve as a {\em bona fide} probability distribution.

I do realize that linearization is a crude and questionable procedure characterized long time ago by Roger Penrose in highly critical terms \cite{rp} ``\dots if we remove the life from Einstein's beautiful theory by steam-rollering it first to flatness and linearity, then we shall learn nothing from attempting to wave the magic wand of quantum theory over the resulting corpse''. Nevertheless, I did make full use of an important feature of the Penrose approach---the correspondence (fully underscored in the spinorial formulation) between the electromagnetic field and the Riemann tensor.

In a recent paper \cite{fd} Freeman Dyson questioned altogether the widely accepted view that the gravitational field is just another field waiting to be quantized. He raised the possibility that ``the gravitational field is a statistical concept like entropy or temperature, only defined for gravitational effects of matter in bulk and not for effects of individual elementary particles''. If this indeed would be so and the gravitons would not exist, my analysis will loose its foundation. However, if you cannot do what you would like, you should like what you can do.

The analysis of the gravitational Wigner functional is greatly simplified if we take full advantage of the analogy between the electromagnetic field tensor $f_{\mu\nu}$ and the Riemann tensor $R_{\mu\nu\lambda\rho}$ in linearized gravity. Despite this close analogy, however, the differences between the electromagnetic and gravitational cases are substantial. The most striking difference is the divergence of the gravitational Wigner functional in the limit of zero temperature which might signify some problems with the ground state. I will show that the source of these problems is found in the singular behavior of Riemann correlators.

In order to make the analogy between the electromagnetic field and the gravitational field unambiguous we need an anchor point that will enable us to convert the loose analogy into precise mathematical formulas. I choose as this anchor point the universal formula for the energy valid for any free field expressed in terms of annihilation and creation operators. For all massless particles this formula can be written in the form:
\begin{eqnarray}\label{anch}
E=\sum_\lambda\int\!\frac{d^3k}{k}\hbar\omega\,a^\dagger_\lambda({\bm k})a_\lambda({\bm k}),
\end{eqnarray}
where $k=|{\bm k}|$ and $\lambda=\pm$ is the sign of helicity. I have chosen the relativistic normalization of the operators so that their commutation relations have the form:
\begin{eqnarray}\label{cr}
\left[a_\lambda({\bm k}),a_{\lambda'}^\dagger({\bm k'})\right]=\delta_{\lambda\lambda'}\,k\,\delta^{(3)}({\bm k}-{\bm k'}).
\end{eqnarray}
This normalization leads to the relativistically invariant volume element in (\ref{anch}) with $k$ in the denominator.

Quantum statistical properties (Planck spectrum, Einstein A and B coefficients, etc.) are, of course, the same for the gas of photons and (linear) gravitons. Both particles are massless bosons with two helicity states. Large differences appear only in the description in terms of fields localized in space.

The analogy between electromagnetism and gravity will be build on the universal formula (\ref{anch}) but the details will be most clearly seen in the spinorial formalism of relativity theory \cite{pr}. In this formalism only the true degrees of freedom will appear and there is no need for any gauge-fixing conditions.

In the next section I briefly summarize the properties of this formalism needed for my analysis. In Sections \ref{max} and \ref{lin} I describe the quantized electromagnetic and the linearized gravitational field within the spinorial formalism. In Sections \ref{wig1} and \ref{wig2} I evaluate the Wigner functionals for the thermal states of the electromagnetic and gravitational fields and I discuss the physical properties of these states. In the last Section I discuss the problems with the ground state of the quantized gravitational field. In \ref{aa} I prove an important equation that relates the number of particles operators to the spinorial field operators. Finally, in \ref{bb} I use the Bel-Robinson tensor to fix the numerical coefficient in the normalization factor for the gravitational field.

\section{Spinorial formalism}

The most convenient representation of the fields describing massless particles is in terms of {\em symmetric} spinors. The connection between the spinors and the electromagnetic field tensor and the Riemann tensor will be given in Sec. \ref{max} and \ref{lin}, respectively. Spinor indices will be denoted by capital letters and those for conjugate spinors by dotted letters. We shall need the following $2\times 2$ matrices:
\numparts\label{sp}
\begin{eqnarray}
\left\{g^{\mu{\dot A}B}\right\}&=\left\{I,{\bm\sigma}\right\}^{{\dot A}B},\quad\epsilon^{AB}
=\left(\begin{array}{cc}0&1\\-1&0\end{array}\right)=\epsilon_{AB},\\
S^{\mu\nu AB}&=\textstyle{\frac{1}{2}}\left(g^{\mu{\dot C}A}\epsilon_{{\dot C}{\dot D}}g^{\nu{\dot D}B}-
g^{\nu{\dot C}A}\epsilon_{{\dot C}{\dot D}}g^{\mu{\dot D}B}\right),
\end{eqnarray}
\endnumparts
where ${\sigma}$'s are the Pauli matrices. My spinor conventions are those of Ref.~\cite{cors}. They differ slightly from the conventions of Ref.~\cite{pr}. The spinor indices take on the values $(0,1)$ and they are raised and lowered as follows:
\begin{eqnarray}
\phi^A&=\epsilon^{AB}\phi_B,\quad \phi_B=\phi^A\epsilon_{AB}.
\end{eqnarray}
The spinorial wave function $\phi_{AB\dots L}(x)$ obeys the wave equation of the same form for all massless particles \cite{pr},
\begin{eqnarray}\label{fe}
g^{\mu{\dot A}A}\partial_\mu\phi_{AB\dots L}(x)=0.
\end{eqnarray}
The number of indices is equal to $2H$; twice the absolute value of the helicity.

The Fourier representation of the general solution of this equation, \begin{eqnarray}\label{sup}
\fl\qquad{\phi}_{AB\dots L}(x)
=\int\!\frac{d^3k}{(2\pi)^{3/2}k}\kappa_A\kappa_B\dots\kappa_L
\left[f_+({\bm k})e^{-ik\cdot x}+f_-^*({\bm k})e^{ik\cdot x}\right],
\end{eqnarray}
expresses the decomposition of the field into harmonic oscillators.  The presence of the volume element $d^3k/k$ underscores the relativistic content of this formula. The spinors $\kappa_A$ are related to the integration variables ${\bm k}$ through the formulae:
\begin{eqnarray}\label{f1}
\kappa_{\dot{A}}g^{\mu{\dot A}A}\kappa_A=k^\mu,\quad \kappa^{\dot A}\kappa^A=\textstyle\frac{1}{2}g^{\mu{\dot A}A}k_\mu.
\end{eqnarray}
The equations (\ref{f1}) do not determine the overall phase of $\kappa_A$. However, this phase is not significant because it can be absorbed by a change of the phases of the amplitudes $f_\pm(\bm k)$ in classical theory or the phases of the annihilation and creation operators in quantum theory. A convenient choice of the spinor $\kappa_A$ is:
\begin{eqnarray}\label{eta}
\{\kappa_A\}=\frac{1}{\sqrt{2(k-k_z)}}\left\{k_x-ik_y,k-k_z\right\}.
\end{eqnarray}
The wave equation (\ref{fe}) is satisfied due to the relations:
\begin{eqnarray}\label{id}
g^{\mu{\dot A}A}k_\mu\kappa_A=g^{\mu{\dot A}A}\kappa_{\dot{B}}g_\mu^{{\dot B}B}\kappa_B\kappa_A=2\epsilon^{{\dot A}{\dot B}}\kappa_{\dot{B}}\epsilon^{AB}\kappa_A\kappa_B=0.
\end{eqnarray}

The field equations retain the same form (\ref{fe}) also for the field operators ${\hat\phi}_{AB\dots L}(x)$ in quantum field theory. In this case the amplitudes $f_\pm(\bm k)$ in the expansion (\ref{sup}) into plane waves must be replaced by the annihilation and creation operators of particles with positive and negative helicity,
\begin{eqnarray}\label{sup1}
\fl\qquad{\hat\phi}_{AB\dots L}(x)
=\gamma\int\!\frac{d^3k}{(2\pi)^{3/2}k}\kappa_A\kappa_B\dots\kappa_L
\left[a_+({\bm k})e^{-ik\cdot x}+a_-^\dagger({\bm k})e^{ik\cdot x}\right].
\end{eqnarray}
The prefactor $\gamma$ is essential because the field operator ${\hat\phi}_{AB\dots L}(x)$ carries the dimensionality of the corresponding physical field while the dimensionality of the integral on the right-hand side is {\em fixed} by the canonical commutation relations (\ref{cr}). It follows from these relations that the annihilation and creation operators have the dimension of length. Therefore, the dimension of the integral is $1/{\rm length}^{H+1}$.

In what follows I shall need the following equality proved in \ref{aa}:
\begin{eqnarray}\label{form}
\fl\qquad{\int\hspace{-0.5cm}\sum}\!h({\bm k})a^\dagger_\lambda({\bm k})a_\lambda({\bm k})&=\int\!d^3r\int\!d^3r'\,{\tilde h}({\bm r}-{\bm r}')\nonumber\\
&\times{\hat\phi}_{\dot{A}\dot{B}\dots\dot{L}}({\bm r},0)g^{0\dot{A}A}g^{0\dot{B}B}\!\dots g^{0\dot{L}L}{\hat\phi}_{AB\dots L}({\bm r'},0),
\end{eqnarray}
where the symbol ${\int\hspace{-0.3cm}\Sigma}$ stands for $\Sigma_\lambda\!\int\!d^3k/k$, the function $h({\bm k})$ must be invariant under reflections, $h({\bm k})=h(-{\bm k})$, and
\begin{eqnarray}\label{ht}
{\tilde h}({\bm r})=\int\!\frac{d^3k}{(2\pi)^3}\,e^{i{\bm k}\cdot{\bm r}}\frac{h({\bm k})}{\gamma^2k^{2H-1}}.
\end{eqnarray}
For odd functions, $h({\bm k})=-h(-{\bm k})$, the contribution from negative helicity on the left hand side enters with the negative sign. In the formula (\ref{form}) and also in (\ref{emt}), (\ref{energy}), (\ref{energy1}), and (\ref{spr}) normal ordering of annihilation and creation operators is implied.

The general formulation will now be applied to the electromagnetic field and then to the gravitational field. The corresponding annihilation and creation operators will be denoted by $(c,c^\dagger)$ and $(g,g^\dagger)$, respectively, while $(a,a^\dagger)$ will be used in the general case. The well established electromagnetic case will serve as a guide in the construction in the gravitational case. I shall often use interchangeably the classical fields and their quantum counterparts but it should be clear from the context what is meant.

\section{Quantized Maxwell theory}\label{max}

Classical electromagnetic field may be described by the second-rank symmetric spinor $\phi_{AB}(x)$. The corresponding field operator ${\hat\phi}_{AB}(x)$ is connected with the electro\-magnetic field operator ${\hat f}_{\mu\nu}$ through the formulas (dotted indices for operators imply Hermitian conjugation):
\begin{eqnarray}\label{ftophi}
{\hat\phi}_{AB}(x)=\textstyle{\frac{1}{4\sqrt{2}}}
S^{\mu\nu}_{\;\;\;AB}{\hat f}_{\mu\nu}(x),\\
{\hat f}_{\mu\nu}(x)
=\frac{1}{\sqrt{2}}\left(S_{\mu\nu}^{\;\;\;AB}{\hat\phi}_{AB}(x)
+S_{\mu\nu}^{\;\;\;{{\dot A}{\dot B}}}{\hat\phi}_{{\dot A}{\dot B}}(x)\right).
\end{eqnarray}
The numerical coefficients in these formulas follows from the convention adopted in \cite{bb}. The value of the electromagnetic prefactor $\gamma_E$ appearing in the formula:
\begin{eqnarray}\label{sup2}
\fl\qquad{\hat\phi}_{AB}(x)
=\gamma_E\int\!\frac{d^3k}{(2\pi)^{3/2}k}\kappa_A\kappa_B
\left[c_+({\bm k})e^{-ik\cdot x}+c_-^\dagger({\bm k})e^{ik\cdot x}\right],
\end{eqnarray}
can be found by comparing the expression (\ref{anch}) for the energy operator of the electromagnetic field with the one constructed from the 00-component of the energy-momentum tensor. In the spinorial representation this tensor has the form:
\begin{eqnarray}\label{emt}
{\hat T}^{\mu\nu}=2\epsilon_0{\hat\phi}_{{\dot A}{\dot B}}g^{\mu{\dot A}A}g^{\nu{\dot B}B}{\hat\phi}_{AB},
\end{eqnarray}
where $\epsilon_0$ is the permittivity of free space. Thus, the following two expressions must be equal:
\begin{eqnarray}\label{energy}
{\int\hspace{-0.5cm}\sum}\hbar\omega\,c^\dagger_\lambda({\bf k})c_\lambda({\bm k})=2\epsilon_0\!\int\!d^3r\,{\hat\phi}_{{\dot A}{\dot B}}g^{0{\dot A}A}g^{0{\dot B}B}{\hat\phi}_{AB}.
\end{eqnarray}
The left hand side with the use of (\ref{form}) can be expressed in terms of spinors as follows:
\begin{eqnarray}\label{energy1}
{\int\hspace{-0.5cm}\sum}\hbar\omega\,c^\dagger_\lambda({\bf k})c_\lambda({\bm k})=\frac{\hbar c}{\gamma_E^2}\int\!d^3r\,{\hat\phi}_{{\dot A}{\dot B}}g^{0{\dot A}A}g^{0{\dot B}B}{\hat\phi}_{AB}.
\end{eqnarray}
Comparing this result with (\ref{energy}) we must choose
$\gamma_E=\sqrt{\hbar c/2\epsilon_0}$ and we obtain finally:
\begin{eqnarray}\label{fem}
\fl\qquad{\hat\phi}_{AB}(x)=\sqrt{\hbar c/2\epsilon_0}\int\!\frac{d^3k}{(2\pi)^{3/2}k}\kappa_A\kappa_B
\left[c_+({\bm k})e^{-ik\cdot x}+c_-^\dagger({\bm k})e^{ik\cdot x}\right].
\end{eqnarray}

\section{Quantized linearized gravity}\label{lin}

In the standard approach to linearized gravity (see, for example, \cite{mtw}) one starts from the decomposition of the metric tensor into the background metric (usually Minkowskian) and a small addition, $g_{\mu\nu}=\eta_{\mu\nu}+h_{\mu\nu}$. Note that the smallness of $h_{\mu\nu}$ does not necessarily imply the smallness of its derivatives so that the linearized Riemann tensor does not have to be small. Next, one proceeds to express the linearized Riemann tensor (I shall keep referring to this tensor as the Riemann tensor even though in this case it effectively reduces to the Weyl tensor) in terms of $h_{\mu\nu}$ and its derivatives. I will bypass all these intermediate steps and following \cite{pr} I will connect directly $R_{\mu\nu\lambda\rho}$ to its representation by the symmetric fourth-rank spinor,
\begin{eqnarray}
\phi_{ABCD}=\textstyle{\frac{1}{16}}
S^{\mu\nu}_{\;\;\;AB}S^{\lambda\rho}_{\;\;\;CD}R_{\mu\nu\lambda\rho},\label{rela}\\
R_{\mu\nu\lambda\rho}=\frac{1}{4}\left(S_{\mu\nu}^{\;\;\;AB}S_{\lambda\rho}^{\;\;\;CD}\phi_{ABCD}
+S_{\mu\nu}^{\;\;\;{\dot A}{\dot B}}S_{\lambda\rho}^{\;\;\;{\dot C}{\dot D}}\phi_{{\dot A}{\dot B}{\dot C}{\dot D}}\right).\label{relb}
\end{eqnarray}
In order to express the field operator $\hat{R}_{\mu\nu\lambda\rho}$ in terms of annihilation and creation operators of gravitons we need the normalization factor $\gamma_G$ in the formula:
\begin{eqnarray}\label{fg}
\fl\qquad{\hat\phi}_{ABCD}(x)=\gamma_G\int\!\frac{d^3k}{(2\pi)^{3/2}k}
\kappa_A\kappa_B\kappa_C\kappa_D
\left[g_+({\bm k})e^{-ik\cdot x}+g_-^\dagger({\bm k}) e^{ik\cdot x}\right].
\end{eqnarray}
The calculation of $\gamma_G$ is a nontrivial task since the expression for the energy of the gravitational field in terms of the Riemann tensor is yet to be determined. Therefore, at this stage I will use a very good substitute --- the Bel-Robinson tensor. The time component of this tensor $T^{0000}$ is positive and it may play sometimes the role of the energy. It has even been called the {\em superenergy} \cite{bel,bel1,bel2}. As is shown in \ref{bb}, the calculation of $\gamma_G$ in this way gives:
\begin{eqnarray}\label{gam}
\gamma_G=\sqrt{8\pi}\,\ell_P,
\end{eqnarray}
where $\ell_P=\sqrt{\hbar G/c^3}$ is the Planck length. Of course, from pure dimensional analysis we can deduce that $\gamma_G$ has the dimension of length so that it must be proportional to the Planck length but the determination of the numerical coefficient requires the calculations in \ref{bb}. Having determined $\gamma_G$ we may freely pass from the description in terms of oscillators labeled by ${\bm k}$ and $\lambda$ to the description in terms fields living in space-time with the use of the formula:
\begin{eqnarray}\label{fg1}
\fl\quad{\hat\phi}_{ABCD}(x)=\sqrt{8\pi\hbar G/c^3}\int\!\frac{d^3k}{(2\pi)^{3/2}k}
\kappa_A\kappa_B\kappa_C\kappa_D
\left[g_+({\bm k})e^{-ik\cdot x}+g_-^\dagger({\bm k}) e^{ik\cdot x}\right].
\end{eqnarray}

\section{Wigner functional of the electromagnetic field}\label{wig1}

The main tool in the study of the fluctuations will be here the Wigner functional at finite temperature. This functional for the electromagnetic field was obtained in \cite{ibb} but I shall sketch here the main steps of the derivation to continue easily to the gravitational case. I start with the Wigner function at finite temperature for the one-dimensional harmonic oscillator \cite{davies,hosw},
\begin{eqnarray}\label{wt}
W_T(x,p)=C\exp\left[-2\tanh\!\left(\!\frac{\hbar\omega}{2k_BT}\!\right)
\frac{H(p,x)}{\hbar\omega}\right],
\end{eqnarray}
where $H(p,x)=p^2/2m+m\omega^2x^2/2$ is the Hamiltonian of the harmonic oscillator. The ratio $N=H(p,x)/\hbar\omega$ is the number of quanta (the energy divided by the energy of one quantum) expressed in terms of classical variables $(p,x)$. At $T=0$, i.e. in the ground state, $W_G=C\exp(-2N)$. The normalization constant $C$ is unimportant since for the infinite number of oscillators only the relative probabilities can be determined.

The Wigner functional of the thermal state of the electro\-magnetic field is constructed by replacing a single oscillator with the whole collection of oscillators labeled by $\bm k$ and $\lambda$ \cite{ibb},
\begin{eqnarray}\label{wigel}
\fl W_{EM}^T[{\bm E},{\bm B}]=\exp\left[-\int\!\!d^3r\!\!\int\!\!d^3r'f_E(\vert{\bm r}-{\bm r}\,'\vert)\left({\bm E}({\bm r}\,)
 \!\cdot\!{\bm E}({\bm r}\,')+c^2{\bm B}({\bm r}\,)\!\cdot\!{\bm B}({\bm r}\,')\right)\right].
\end{eqnarray}
The electromagnetic correlation function $f_E(r)$ is the three-dimensional Fourier transform of the function appearing in the Wigner function for the one-dimensional oscillator,
\begin{eqnarray}\label{kerel}
\fl\qquad f_E(r)=\frac{1}{2\gamma_E^2}\int\!\frac{d^3k}{(2\pi)^3}\frac{\tanh
 \left(\ell_Qk/2\right)}{k}e^{i{\bm k}\cdot{\bm r}}
=\frac{\epsilon_0}{2\pi\hbar c\ell_Q r\sinh(\pi r/\ell_Q)},
\end{eqnarray}
where the quantum thermal length is $\ell_Q=\hbar c/k_BT$=0.0023m/T[K].
In the derivation of (\ref{wigel}) I used (\ref{form}) and the following relation:
\begin{eqnarray}\label{rel1}
\fl\qquad\phi_{\dot{A}\dot{B}}({\bm r})g^{0\dot{A}A}g^{0\dot{B}B}\phi_{AB}({\bm r'})=\frac{1}{4}\left({\bm E}({\bm r})\!\cdot\!{\bm E}({\bm r}\,')+c^2{\bm B}({\bm r})\!\cdot\!{\bm B}({\bm r}\,')\right).
\end{eqnarray}
From the integral representation (\ref{kerel}) of $f_E(r)$ we obtain in the classical limit,
\begin{eqnarray}\label{lim}
\lim_{\hbar\to 0}f_E(r)=\frac{\epsilon_0}{2}\,\delta(\bm r).
\end{eqnarray}
Therefore, in this limit the Wigner functional becomes the Boltzmann distribution,
\begin{eqnarray}\label{boltz}
W_{EM}^{\rm cl}=\exp(-E_{EM}/k_BT).
\end{eqnarray}
At the other end, at $T=0$, the Wigner functional is equal to:
\begin{eqnarray}\label{zeld}
\fl\qquad W_{EM}^0[{\bm E},{\bm B}]=\exp\left[-\epsilon_0\int\!\!d^3r\!\!\int\!\!d^3r'\frac{{\bm E}({\bm r}\,)
 \!\cdot\!{\bm E}({\bm r}\,')+c^2{\bm B}({\bm r}\,)\!\cdot\!{\bm B}({\bm r}\,')}{2\pi^2\hbar c\vert{\bm r}-{\bm r}\,'\vert^2}\right],
\end{eqnarray}
i.e. equal to $\exp(-2N)$ where $N$ is the total number of photons as calculated by Zeldovich \cite{zeld,ibb}.

\section{Wigner functional of the gravitational field}\label{wig2}

The thermal Wigner functional of the gravitational field can be constructed in the same way as for the electromagnetic field. The resulting formula is:
\begin{eqnarray}\label{wigg}
\fl W_G^T(R)=\exp\left[- \int\!\!d^3r\!\!\int\!\!d^3r'f_G(\vert{\bm r}-{\bm r}\,'\vert)\sum\limits_{ij}\left({\cal E}_{ij}({\bm r}\,)
 {\cal E}_{ij}({\bm r}\,')+{\cal B}_{ij}({\bm r}\,){\cal B}_{ij}({\bm r}\,')\right)\right],
\end{eqnarray}
where ${\cal E}_{ij}=R_{i0j0}$ and ${\cal B}_{ij}=\textstyle{\frac{1}{2}}\epsilon_{ikl}R_{j0}^{\;\;\;kl}$ are the so called electric and magnetic parts of the curvature tensor \cite{mb,kt}. In the derivation of this formula I used the following gravitational counterpart of (\ref{rel1}) derived with the help of (\ref{form}):
\begin{eqnarray}\label{rep}
\fl\phi_{\dot{A}\dot{B}\dot{C}\dot{D}}({\bm r})g^{0\dot{A}A}g^{0\dot{B}B}g^{0\dot{C}C} g^{0\dot{D}D}
\phi_{ABCD}({\bm r'})=\sum_{ij}\left({\cal E}_{ij}({\bm r}){\cal E}_{ij}({\bm r'})+{\cal B}_{ij}({\bm r}){\cal B}_{ij}({\bm r'})\right).
\end{eqnarray}
The gravitational correlation function $f_G(r)$ can be explicitly evaluated and is given by the following counterpart of the formula (\ref{kerel}),
\begin{eqnarray}\label{kerg}
\fl\qquad\qquad f_G(r)&=\frac{1}{\gamma_G^2}\int\!\frac{d^3k}{(2\pi)^3}\frac{2\tanh \left(\ell_Qk/2\right)}{k^3}e^{i{\bm k}\cdot{\bm r}}\nonumber\\
 &=\frac{1}{8\pi^4\ell_Gr}\left[\frac{\pi^2}{3}+\ln(\zeta)\ln(1+\zeta)
 +{\rm Li}_2(1-\zeta)+{\rm Li}_2(-\zeta)\right],
\end{eqnarray}
where $\ell_G=Gk_BT/c^4=1.14\times 10^{-67}$mT[K] is the gravitational thermal length, $\zeta=\coth(\pi r/2\ell_Q)$, and ${\rm Li}_2$ is the dilogarithm function.

There is a great similarity between the probability distributions of various field configurations for the electromagnetic field and the gravitational field. In both cases the smaller the distance between the points, the more likely it is that the electromagnetic field or the curvature tensor at these points will have opposite signs. Thus, the formula (\ref{wigg}) may be viewed as a realization of the Wheeler concept of the virtual gravitational foam \cite{jaw}. Of course, in the linearized version of gravity there is no room for fluctuations of topology envisaged by Wheeler but the rapid changes of geometry at small distances do show a certain similarity with Wheeler's foam.

In the classical limit in the electromagnetic case we obtained the standard Boltzmann distribution (\ref{boltz}) and the same result holds in the gravitational case, $W_G^{\rm cl}(R)=\exp(-E_G/k_BT)$, where
\begin{eqnarray}\label{ham}
E_G= c^4\int\!\!d^3r\!\!\int\!\!d^3r'
\sum\limits_{ij}\frac{{\cal E}_{ij}({\bm r}\,)
 {\cal E}_{ij}({\bm r}\,')+{\cal B}_{ij}({\bm r}\,){\cal B}_{ij}({\bm r}\,')}{32\pi^2G\vert{\bm r}-{\bm r}\,'\vert}.
\end{eqnarray}
One may check that $E_G$ is indeed equal to (\ref{anch}) by expressing (\ref{ham}), with the use of (\ref{form}) and (\ref{rep}), in terms of annihilation and creation operators,
\begin{eqnarray}\label{genergy}
E_G=\int\hspace{-0.5cm}\sum\hbar\omega\,g^\dagger_\lambda({\bf k})g_\lambda({\bm k}).
\end{eqnarray}
The equality of these two expressions proves the self-consistency of the whole procedure. The classical (high-temperature) limit gives the correct expression for the energy both in the electromagnetic and in the gravitational case. The nonlocal form of $E_G$ fully confirms the belief that there is no ``local gravitational energy-momentum'' \cite{mtw}.

There are, however, substantial differences between the gravitational and electro\-magnetic correlation functions. At large distances the gravitational correlation function $f_G(r)$ does not fall-off exponentially, as in electromagnetism, but has a {\em long tail} equal to its classical limit $f_G(r)\approx f^{\rm cl}_G(r)= 1/(32\pi^2\ell_Gr)$. There is also a striking difference in the behavior at $T\to 0$, as described in the next Section.

\section{Problems with the ground state}\label{grnd}

A puzzling phenomenon is the logarithmic divergence of the gravitational correlation function $f_G(r)$ in the limit, when $T\to 0$. This fact indicates that there might be a problem with the gravitational ground state.

The ground state of the quantized gravitational field has been studied before by Kucha{\v{r}} \cite{kk} and Hartle \cite{jbh}. They arrived at a formula for the ground state functional but their analogy between the electromagnetism and gravity is based on a different paradigm. They saw the analogy between the electromagnetic potentials and the metric tensor while in my approach, based on the spinorial description, there is a correspondence between the electromagnetic field tensor and the Riemann tensor.

The roots of the problem are in the expression (\ref{ham}) for the energy. To see the connection between the energy and the ground state let us consider the wave functions of a one-dimensional harmonic oscillator in the position representation and the momentum representation:
\begin{eqnarray}\label{gs}
\psi(x)=N_x\exp\left(-\frac{m\omega x^2}{2\hbar\omega}\right),\quad
{\tilde\psi}(p)=N_p\exp\left(-\frac{p^2}{2m\hbar\omega}\right).
\end{eqnarray}
In the position representation the exponent is equal to the potential energy divided by $\hbar\omega$ and in the momentum representation it is the kinetic energy divided by $\hbar\omega$. Applying the same prescription to the electromagnetic field we obtain the Wheeler expression \cite{jaw} (also rederived in \cite{kk} and \cite{jbh}). This follows directly from the observation that the division by $\hbar\omega$ in quantum mechanics is represented in quantum field theory by the following kernel
\begin{eqnarray}\label{ker}
 \int\!\frac{d^3k}{\hbar\omega_k}e^{i{\bm k}\cdot({\bm r}-{\bm r}')} =
 \frac{4\pi}{\hbar c\vert{\bm r} - {\bm r}'\vert^2}.
\end{eqnarray}
However, in the gravitational case we do not reproduce the Kucha{\v{r}}-Hartle result but we encounter the same logarithmic divergence that appeared in the Wigner functional. The convolution of this kernel with the already present kernel $1/\vert{\bm r} - {\bm r}'\vert$ in the formula (\ref{energy}) for the energy leads to the logarithmic divergence at large values of $r$. The same divergence appears already in the formula for the number of gravitons obtained by setting $h({\bm k})=1$ and $H=2$ in (\ref{ht}).

The divergence of the gravitational probability at $T=0$ can also be related to the behavior of the two-point function (sometimes called Riemann correlator \cite{frv}) at small distances. With the use of the formulas (\ref{sup1}) and (\ref{relb}) we can evaluate the vacuum expectation value
\begin{eqnarray}\label{rc}
\langle 0|{\hat\phi}_{AB\dots L}(x){\hat\phi}_{{\dot{A}}{\dot{B}}\dots{\dot{L}}}(\bm x')|0\rangle
\end{eqnarray}
for the spinor fields of rank $2H$. The resulting expression will have $2H$ derivatives acting on $\Delta^{(+)}(x-x')$---the invariant (positive frequency) solution of the wave equation. For a Gaussian distribution there is a simple relation between the matrix of the second moments $\langle x_ix_j\rangle$ and the matrix in the probability distribution. Namely, one is the inverse of the other. There exists an analogy between this discrete case and our field-theoretic continuous case. The two-point function is the analog of the covariance matrix and the correlation functions $f_E$ and $f_G$ in (\ref{wigel}) and (\ref{wigg}) are the analogs of the inverse of the covariance matrix. The divergence of $f_G$ at zero temperature can be traced to the nonexistence of the inverse of the gravitational two-point function. The presence of four derivatives in this function results in the fourth power of $k$ in the Fourier transform and its inverse is not integrable at small $k$. In the electromagnetic case we have the second derivatives and this leads to an integrable Fourier transform.

A technical cure for the problem with non-integrability could be to disallow the interchange of the integrations over ${\bm r}$ and ${\bm r}'$ in the formula (\ref{wigg}) and the evaluation of the limit when $T\to 0$. Unfortunately, the limit would then depend on the behavior of the Riemann tensor at large distances. Since the divergence is of the ``infrared type'' it can presumably be cured by assuming a finite radius of the Universe but this is outside the scope of this paper. The problem with the ground state might also mean, however, that there is some truth in Dyson's hypothesis.

\section{Conclusions}

I have shown that the quantization of the linearized gravitational field that employs only the Riemann tensor (and not the metric tensor) can be achieved without any reference to the canonical formalism. In this approach the complications arising in the process of extracting true degrees of freedom never appear. The main result is the evaluation of the probability distribution of various geometries in the thermal state of the gravitational field. An encountered puzzle is the divergence of this probability distribution at $T\to 0$ that might signify some problems with the ground state.

\ack

I would like to thank Zofia Bialynicka-Birula for very fruitful criticism and Albert Roura and Jerzy Kijowski for discussions and references. This research was financed by the Polish National Science Center Grant No. 2012/07/B/ST1/03347.

\appendix

\section{Proof of the equality (\ref{form})}\label{aa}

This equality can be proven by first inverting the Fourier transform in (\ref{sup1}) at $t=0$ which leads to:
\begin{eqnarray}\label{sup3}
\fl\qquad\kappa_A\kappa_B\dots\kappa_La_+({\bm k})
+{\tilde\kappa}_A{\tilde\kappa}_B\dots{\tilde\kappa}_La_-^\dagger(-{\bm k})=\frac{k}{\gamma}\int\!\frac{d^3r}{(2\pi)^{3/2}}e^{-i{\bm k}\cdot{\bm r}}{\hat\phi}_{AB\dots L}({\bm r}),
\end{eqnarray}
where ${\tilde\kappa}_A({\bm k})=\kappa_A(-{\bm k})$. In the next step I evaluate the product of both sides with their hermitian conjugate inserting the unit matrices $g^{0\,{\dot A}A}$ to enable the summation convention by balancing dotted and undotted indices. Using the relations:
\begin{eqnarray}\label{bal}
\fl\qquad\kappa_{\dot A}g^{0\,{\dot A}A}\kappa_A=k={\tilde\kappa}_{\dot A}g^{0\,{\dot A}A}{\tilde\kappa}_A,\quad \kappa_{\dot A}g^{0\,{\dot A}A}{\tilde\kappa}_A=0={\tilde\kappa}_{\dot A}g^{0\,{\dot A}A}\kappa_A,
\end{eqnarray}
we eliminate the cross terms and obtain:
\begin{eqnarray}\label{spr}
\fl a_+^\dagger({\bm k})a_+({\bm k})+a_-^\dagger(-{\bm k})a_-(-{\bm k})&=\frac{1}{\gamma^2k^{2H-2}}\frac{1}{(2\pi)^{3}}\int\!d^3r\,e^{i{\bm k}\cdot{\bm r}}\int\!d^3r'\,e^{-i{\bm k}\cdot{\bm r}'}\nonumber\\
&\times{\hat\phi}_{\dot{A}\dot{B}\dots\dot{L}}({\bm r},0)g^{0\dot{A}A}g^{0\dot{B}B}\!\dots g^{0\dot{L}L}{\hat\phi}_{AB\dots L}({\bm r'},0).
\end{eqnarray}
Multiplying both sides of this equality by $h(\bm k)$ and integrating  over ${\bm k}$ we obtain the relation (\ref{form}).

\section{Bel-Robinson tensor}\label{bb}

To find the normalization factor $\gamma_G$ I use the Bel-Robinson tensor \cite{pr,bel,bel1,bel2}. This is a fourth-rank tensor $T^{\mu\nu\lambda\rho}$ which to some extent can play the role of the energy-momentum tensor; in the Einstein-Maxwell system it satisfies the continuity equation $\partial_\rho( T^{\mu\nu\lambda\rho}_{EM}+T^{\mu\nu\lambda\rho}_{G})=0$. Since the sum of the contributions from electromagnetism and gravity is {\em conserved}, the normalization of the electromagnetic part determines the normalization of the gravitational part. To evaluate $\gamma_G$ I shall use the spinorial form of the Bel-Robinson tensor \cite{pr},
\begin{eqnarray}\label{br}
&T^{\mu\nu\lambda\rho}_{EM}=\frac{\epsilon_0}{4}g^{\mu{\dot{A}}A}
g^{\nu{\dot{B}}B}g^{\lambda{\dot{C}}C}g^{\rho{\dot{D}}D}\nonumber\\
&\times\left(3{\rm\bf S}_{ABC}^{\dot{A}\dot{B}\dot{C}}
g^\alpha_{\;C\dot{D}}\partial_\alpha
\phi_{\dot{A}\dot{B}}\,g^\beta_{\;D\dot{C}}\partial_\beta\phi_{AB}
-g^\alpha_{\;D\dot{C}}\partial_\alpha\phi_{\dot{A}\dot{B}}\,
g^\beta_{\;C\dot{D}}\partial_\beta\phi_{AB}\right),\\
&T^{\mu\nu\lambda\rho}_G=\frac{c^4}{16\pi G}g^{\mu{\dot{A}}A}g^{\nu{\dot{B}}B}g^{\lambda{\dot{C}}C}g^{\rho{\dot{D}}D}
\phi_{\dot{A}\dot{B}\dot{C}\dot{D}}\phi_{ABCD},
\end{eqnarray}
where {\bf S} means the symmetrization with respect to the listed indices. The electro\-magnetic and the gravitational contributions to the super-energy can be evaluated with the use of (\ref{fem}) and (\ref{fg}),
\begin{eqnarray}\label{embr1}
\int\!d^3r\,T_{EM}^{0000}=\frac{1}{2}\int\hspace{-0.5cm}\sum\hbar\omega\,k^3 c^\dagger_\lambda({\bf k})c_\lambda({\bm k}),
\end{eqnarray}
\begin{eqnarray}\label{gbr1}
\int\!d^3r\,T_G^{0000}=\frac{\gamma_G^2\;c^3}{16\pi\hbar G}\int\hspace{-0.5cm}\sum\hbar\omega\, k^3 g^\dagger_\lambda({\bf k})g_\lambda({\bm k}).
\end{eqnarray}
These contributions must have the same form since they both represent the same physical quantity. Therefore, $\gamma_G=\sqrt{8\pi}\,\ell_P$, where $\ell_P=\sqrt{\hbar G/c^3}$ is the Planck length.

\end{document}